\documentclass[aps,letterpaper,twocolumn,showpacs,superscriptaddress,amsmath,amsfonts,amssymb,preprintnumbers,showkeys]{revtex4}
\pdfoutput=1
\usepackage[T1]{fontenc}\usepackage[latin1]{inputenc}
\usepackage{dcolumn,graphicx,color,hvfloat,booktabs,microtype,afterpage}

\usepackage[colorlinks,plainpages=false,linkcolor=blue,urlcolor=blue,citecolor=black,pdfpagemode=UseNone,pdfstartview=FitBH]{hyperref}

\newcommand{\La}{LaFeAsO$_{1-\mathrm x}$F$_{\mathrm x}$}
\newcommand{\LAO}{LaAlO$_3$}
\newcommand{\Hc}{$H_{\mathrm c 2}$}
\newcommand{\Hcs}{$H_{\mathrm c 2}^{\parallel \mathrm {ab}}$}
\newcommand{\Hcp}{$H_{\mathrm c 2}^{\parallel \mathrm c}$}
\newcommand{\Rc}{$R_{90}\ $}
\newcommand{\Hirr}{$H_{\mathrm{irr}}$}
\newcommand{\Tc}{$T_{\mathrm c}$}
\newcommand{\Jc}{$J_{\mathrm c}$}
\newcommand{\Heff}{$H_{\mathrm {eff}}$}
\newcommand{\Hirrp}{$H_{\mathrm {irr}}^{\parallel \mathrm c}$}

\begin{document}

\title{Critical current scaling and anisotropy in oxypnictide superconductors}

\author{M.~Kidszun}
\affiliation{IFW Dresden, Institute for Metallic Materials, P.\,O.\,Box 270116, D-01171 Dresden, Germany.}
\affiliation{TU Dresden, Institut f\"{u}r Festkörperphysik, D-01069 Dresden, Germany.}
\author{S.~Haindl}
\affiliation{IFW Dresden, Institute for Metallic Materials, P.\,O.\,Box 270116, D-01171 Dresden, Germany.}

\author{T.~Thersleff}
\affiliation{IFW Dresden, Institute for Metallic Materials, P.\,O.\,Box 270116, D-01171 Dresden, Germany.}

\author{J.~H\"{a}nisch}
\affiliation{IFW Dresden, Institute for Metallic Materials, P.\,O.\,Box 270116, D-01171 Dresden, Germany.}

\author{A.~Kauffmann}
\affiliation{IFW Dresden, Institute for Metallic Materials, P.\,O.\,Box 270116, D-01171 Dresden, Germany.}

\author{K.~Iida}
\affiliation{IFW Dresden, Institute for Metallic Materials, P.\,O.\,Box 270116, D-01171 Dresden, Germany.}

\author{J.~Freudenberger}
\affiliation{IFW Dresden, Institute for Metallic Materials, P.\,O.\,Box 270116, D-01171 Dresden, Germany.}

\author{L.~Schultz}
\affiliation{IFW Dresden, Institute for Metallic Materials, P.\,O.\,Box 270116, D-01171 Dresden, Germany.}
\affiliation{TU Dresden, Institut f\"{u}r Festkörperphysik, D-01069 Dresden, Germany.}

\author{B.~Holzapfel}
\affiliation{IFW Dresden, Institute for Metallic Materials, P.\,O.\,Box 270116, D-01171 Dresden, Germany.}
\affiliation{TU Dresden, Institut f\"{u}r Festkörperphysik, D-01069 Dresden, Germany.}

\begin{abstract}
 
Investigating the anisotropy of superconductors permits an access to fundamental properties. Having succeeded in the fabrication of epitaxial superconducting \La\ thin films we performed an extensive study of electrical transport properties. In face of multiband superconductivity we can demonstrate that a Blatter scaling of the angular dependent critical current densities can be adopted, although being originally developed for single band superconductors. In contrast to single band superconductors the mass anisotropy of \La\ is temperature dependent. A very steep increase of the upper critical field and the irreversibility field can be observed at temperatures below 6\,K, indicating that the band with the smaller gap is in the dirty limit. This temperature dependence can be theoretically described by two dominating bands responsible for superconductivity. A pinning force scaling provides insight into the prevalent pinning mechanism and can be specified in terms of the Kramer model. 

\end{abstract}

\pacs{\vspace{-0.2em}74.25.Sv 74.70.Xa 74.25.Wx 74.62.-c 68.37.Og 74.25.Dw}


\maketitle

Superconductivity in \La, hereafter La--1111, was discovered in 2008 by Kamihara et. al \cite{Kam08}. However, due to the difficulties in fabricating single crystals their intrinsic electronic properties are still under discussion \cite{Kar09,Yan09,Cru08}. Polycrystalline oxypnictides reveal very high upper critical fields as well as weak link behaviour \cite{Hai10}. Multiband superconductivity is predicted from theoretical calculations \cite{Esch09} supported by experimental results on the upper critical field in polycrystalline La--1111 bulk samples \cite{Hun08} as well as on Sm--1111 and Nd--1111 single crystals \cite{Jar08} and, furthermore, nuclear magnetic resonance experiments \cite{Kaw08}. As two distinct superconducting gaps were found in the oxypnictides \cite{Gon09} we simplify matters by considering only two major bands, created by electrons and holes respectively. Regarding multiband superconductivity the anisotropy is of substantial interest. Calculations of average Fermi velocities in oxypnictides yield a remarkably high ratio ${v_{xx}}/{v_{zz}}=15$ \cite{Sin08} resulting in a large anisotropy of the effective electron mass, $\gamma_{\mathrm m}$. This large anisotropy value still has not been confirmed by experiments. Jia et al. \cite{Jia08} reported an anisotropy of the upper critical field, $\gamma_{H_{\mathrm c 2}}$, in Nd--1111 single crystals of around $5$ near $T_{\mathrm c}$. However, $\gamma_{\mathrm m}$ and $\gamma_{H_{\mathrm c 2}}$ in multiband superconductors can not be considered equal as in the anisotropic Ginzburg Landau theory for single band superconductors \cite{Kog02}. There, the relationship between anisotropies of electron masses, penetration depth, coherence length, and upper critical field can be expressed as $\gamma=\sqrt{\frac{m_c}{m_{ab}}}=\frac{\lambda_c}{\lambda_{ab}}=\frac{\xi_{ab}}{\xi_c}=\frac{H_{\mathrm c 2}^{\parallel \mathrm {ab}}}{H_{\mathrm c 2}^{\parallel \mathrm c}}$. Instead of this single anisotropy, it is necessary to distinguish between two anisotropies, $\gamma_{\lambda}={\lambda_c}/{\lambda_{ab}}$ and $\gamma_{H_{\mathrm c 2}}={H_{\mathrm c 2}^{\parallel \mathrm {ab}}}/{H_{\mathrm c 2}^{\parallel \mathrm c}}$ in order to understand the multiband character of pnictide superconductors \cite{Wey09}.

\begin{figure}[htbp]
	\centering
		\includegraphics[width=1.0\columnwidth]{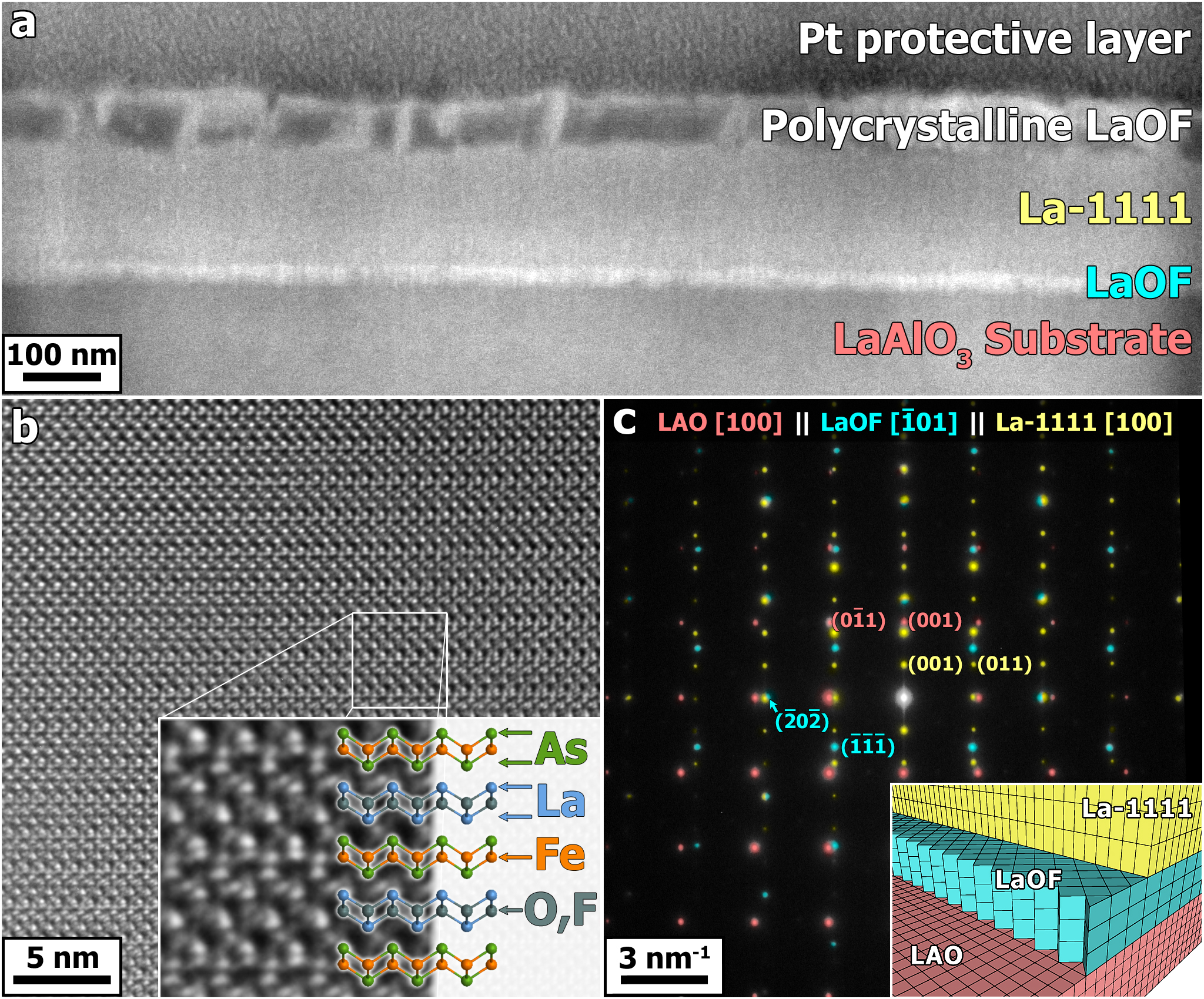}
	\caption{a) \La\ thin film cross section as prepared by FIB. The approximately 150\,nm thick epitaxial portion appears free of secondary phases and grows between two LaOF layers. b) A representative HRTEM micrograph of the epitaxially grown La--1111 phase imaged along the [100] zone axis reveals highly crystalline growth without correlated defects. In the inset enlargement, the positions of the atomic columns are denoted. The image was Fourier filtered for noise reduction. c) The artificially coloured selected area diffraction pattern of the film cross section reveals the epitaxial relationship between \LAO\ substrate, the LaOF intermediate layer, and the La--1111 epitaxial phase.}
	\label{fig:HRTEM}
\end{figure}

The very high upper critical fields off the pnictide superconductors rapidly exceed the capability of common high field measurement systems. Therefore, reliable information about the temperature dependence of \Hc\ and multiband calculations concerning the magnetic phase diagram are difficult to obtain \cite{Jar08}. 
The rather low superconducting transition temperature of the La--1111 system and, accordingly, the low upper critical fields compared to other oxypnictides makes La--1111 thin films good candidates for studying the magnetic phase diagram and theoretical aspects, especially the mass anisotropy. In this work we document the mass anisotropy and its temperature dependence by electrical transport measurements of the critical currents.

Our recent success in fabricating superconducting epitaxially grown \La\ thin films of high crystalline quality \cite{Kid10} opens the way to study their basic properties. The samples were prepared using a standard high vacuum pulsed laser deposition setup at room temperature. An annealing process at about 1000°C realised in evacuated quartz tubes initiates the \La\ phase formation and leads to an epitaxially grown thin film. For a detailed description of the sample preparation please refer to \cite{Bac08}. 
In the following, we focus on the analysis of the critical current density, $J_{\mathrm c}(T,B,\theta)$, and especially its angular dependence. Furthermore, the irreversibility field, $H_{\mathrm{irr}}$, and the prevalent pinning mechanism are discussed. A comprehensive assessment of the film quality confirms textured growth and an absence of phase impurities, ensuring that none of the measurements is affected by grain boundaries or precipitates. Results of standard x--ray diffraction methods are provided in the supplementary. 

\begin{figure}[htbp]
	\centering
		\includegraphics[width=0.71\columnwidth]{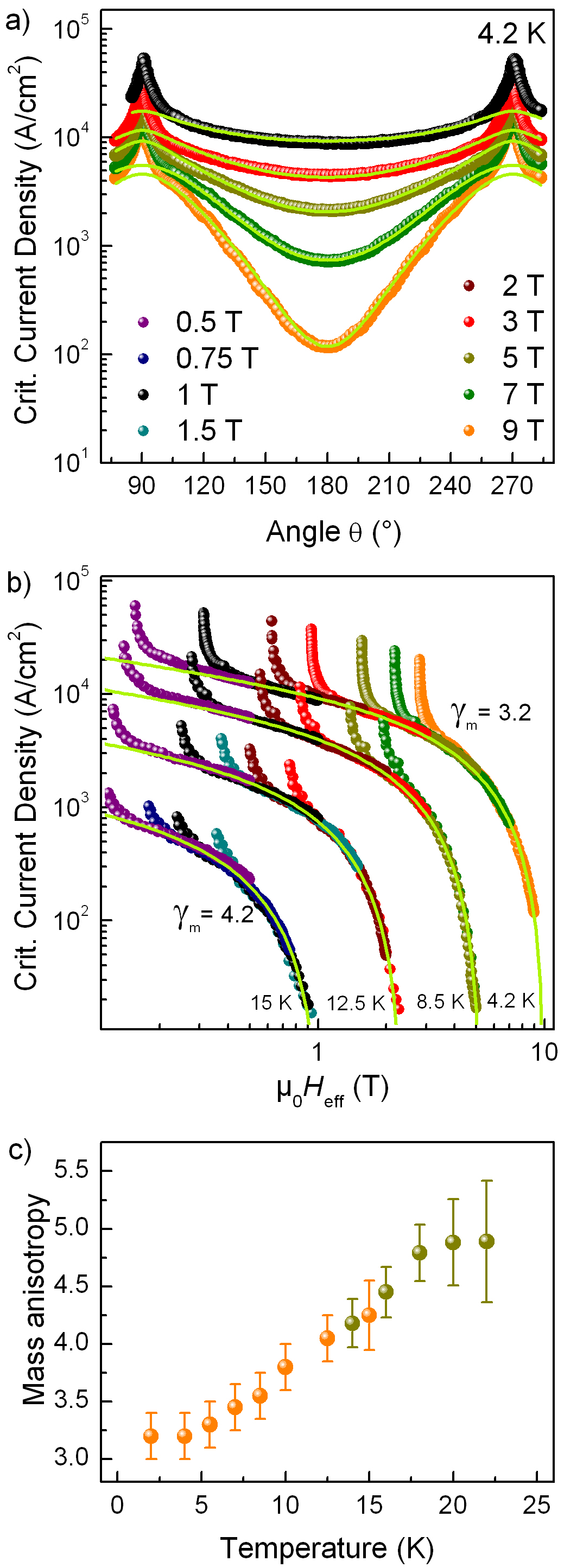}
	\caption{(a) Angular dependence of the critical current density for several magnetic fields at 4\,K. (b) Blatter scaling can be adopted for all temperatures with a temperature dependent anisotropy factor $\gamma$. (c) Temperature dependence of the mass anisotropy determined from Blatter scaling of \Jc\ data at low temperatures and from pulsed field \Hc\ anisotropy data at high temperatures.}
	\label{fig:Criticalcurrents}
\end{figure}

\begin{figure}[htbp]
	\centering
		\includegraphics[width=1\columnwidth]{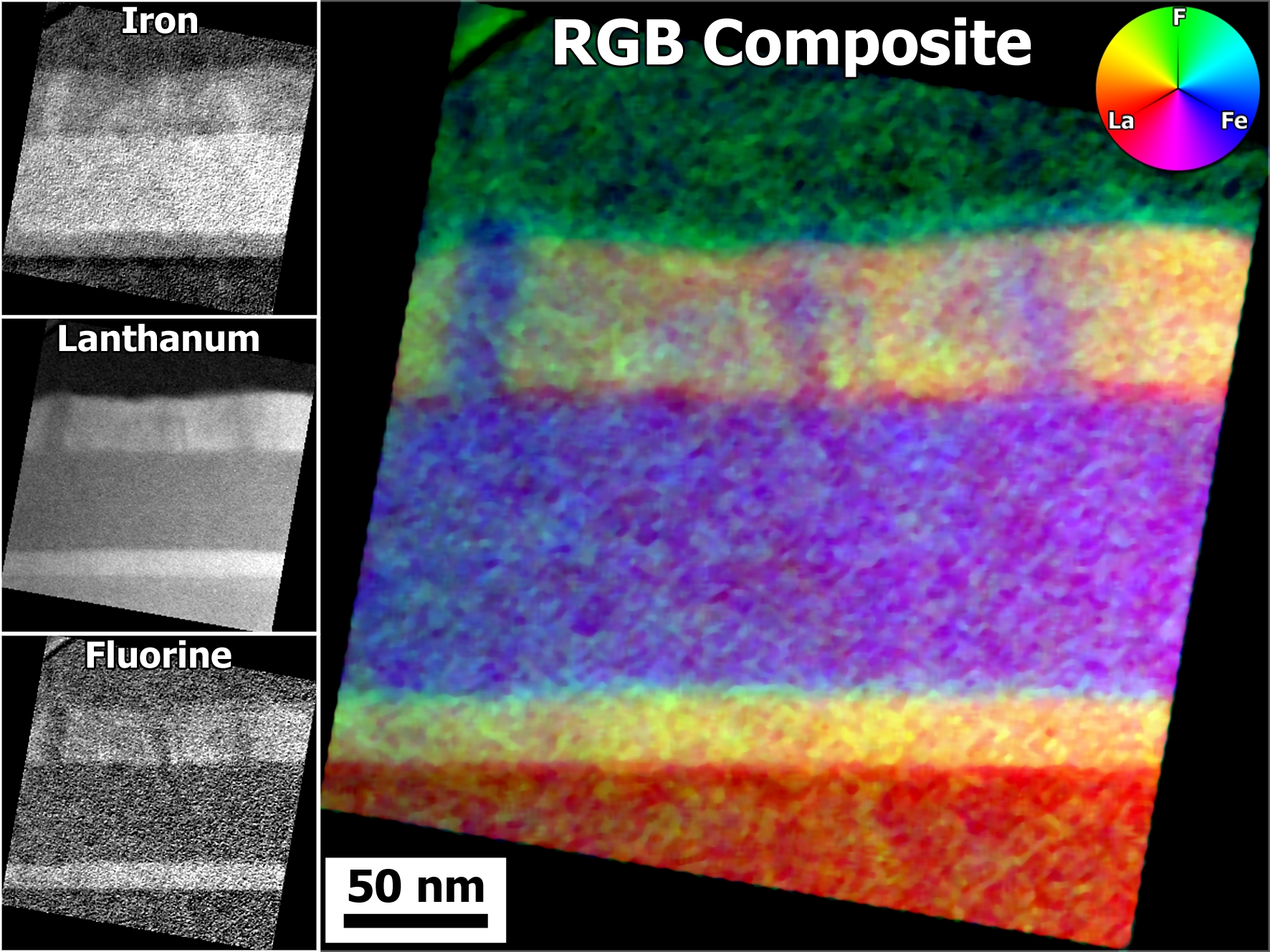}
	\caption{Overview of the La--1111 thin film acquired using the three--window Energy Filtered TEM (EFTEM) imaging technique. The core--loss edges F (K), Fe (L2), and La (M4) were used. The raw data for each element are shown at left and are used to construct the composite RGB image shown in the middle. The epitaxial La--1111 film grows sandwiched between two lanthanum and fluorine--rich layers but is itself chemically homogeneous. As the fluorine baseline is noisy due to a weak signal, it appears as an artifact in the platinum surface layer. 
The lamella was prepared for TEM using the Focused Ion Beam (FIB) in--situ lift--out technique on a Carl Zeiss NVision FIB and was thinned to electron transparency using a 5\,kV Ga$^+$ ion beam. The polished lamella was then investigated in a C$_S$--corrected FEI Titan 80--300 TEM operating at 300\,kV.}
	\label{fig:EFTEM}
\end{figure}

A Scanning Electron Microscopy image of a Focused Ion Beam (FIB) cross section shown in Fig.~\ref{fig:HRTEM}a indicates that the La--1111 phase has a thickness of about 150\,nm and grows sandwiched between two secondary phase layers. These can be identified as an epitaxially--grown LaOF layer between the La--1111 film and the \LAO\ substrate as well as a polycrystalline surface layer (Suppl:~Fig.~\ref{fig:EFTEM}). The La--1111 layer itself is dense and appears to be free of secondary phases, which was subsequently confirmed in the Transmission Electron Microscopy (TEM) investigation. In Fig.~\ref{fig:HRTEM}b, a representative yet very thin region of the La--1111 layer is shown indicating perfect single crystalline growth. The electron beam is parallel to the [100] zone axis revealing the La--1111 atomic structure. The positions of the atomic columns were determined by comparison of the raw data with contrast simulations.  The epitaxial relationship between the various layers is documented by the selected area diffraction pattern in Fig.~\ref{fig:HRTEM}c. The LaOF intermediate layer acts as a buffer for the La--1111 layer to relax its lattice misfit to the \LAO\ substrate.

\begin{figure}[htbp]
	\centering
		\includegraphics[width=1\columnwidth]{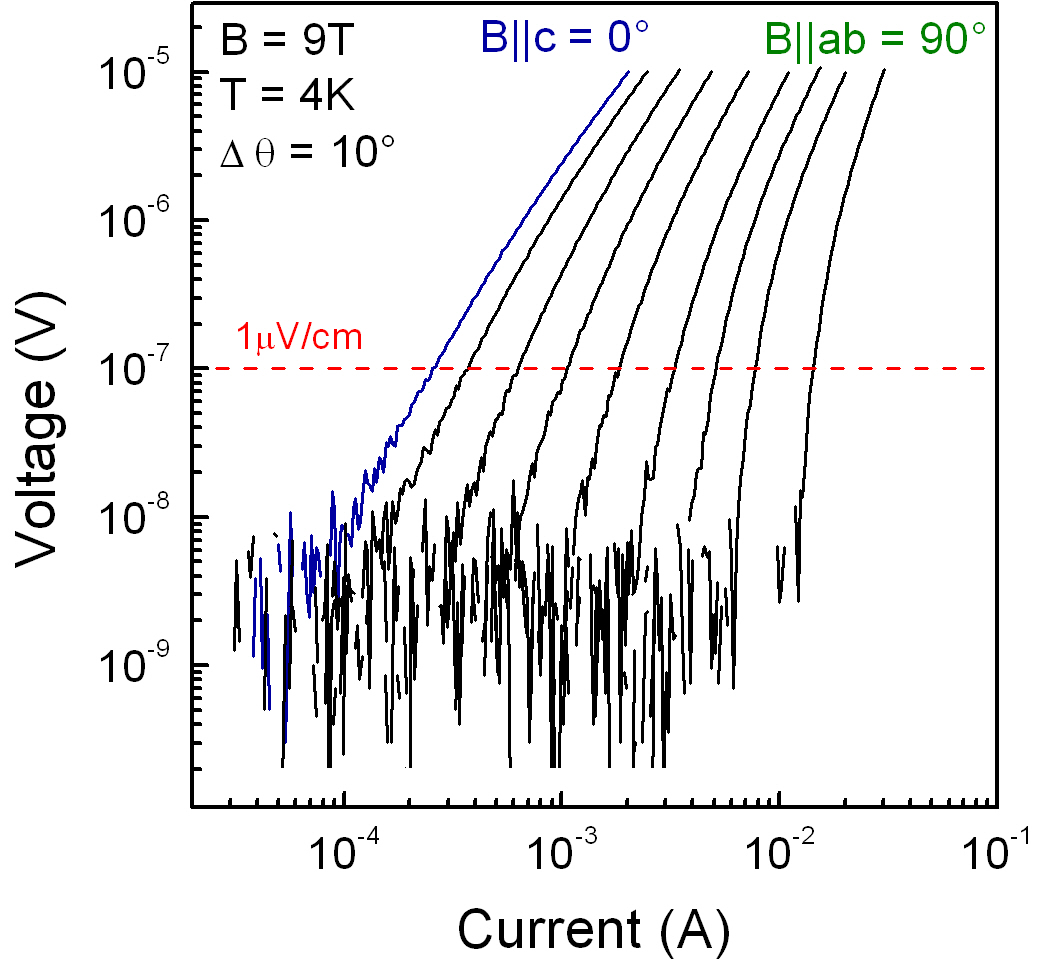}
	\caption{Double logarithmic plot of angular dependent $I(V)$ measurements at 4\,K and 9\,T with an angular step size of 10°. The critical currents were measured in a standard four--point technique on a bridge with 1mm distance between the voltage contacts. For evaluating the critical currents, a criterion E$_{\mathrm{c}} = $1\,$\mu$V/cm was used.}
	\label{fig:IVcurves}
\end{figure}

\begin{figure}[htbp]
	\centering
		\includegraphics[width=1\columnwidth]{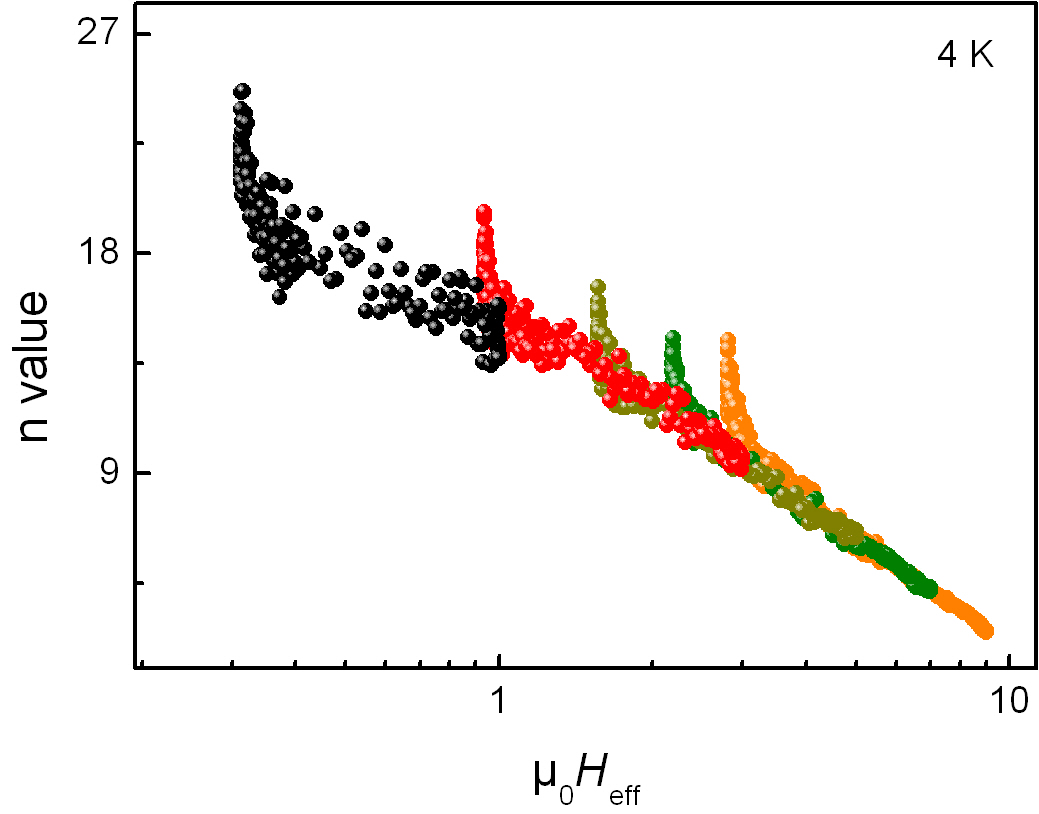}
	\caption{n--value, $U \propto I^n$, in the vicinity of \Jc\ can be scaled with the Blatter approach shown for 4\,K. The entire angular dependent $V(I)$ curves can be scaled and, for this reason, the results are independent of the chosen \Jc\ criterion.}
	\label{fig:nHeps4K}
\end{figure}

Intrinsic properties were investigated by means of electrical transport measurements, which is best accomplished using thin films since transport measurements of the critical current densities on single crystals is very challenging \cite{Mol10}. Critical current densities, $J_{\mathrm c}(T,B,\theta)$, have been measured in standard four point configuration and the applied magnetic field is always aligned perpendicular to the measurement current. A standard \Jc\ criterion of 1$\mu$Vcm$^{-1}$ was applied (Suppl:~Fig.~\ref{fig:IVcurves},\ref{fig:nHeps4K}). At all temperatures, the critical currents have a minimum at $B\parallel \mathrm {c}$ (180°) and a maximum at $B\parallel\mathrm{ab}$ (90° and 270°) as exemplarily shown in Fig.~\ref{fig:Criticalcurrents}a. The maxima at $B\parallel \mathrm {ab}$ can be explained by intrinsic pinning because the coherence length ($\xi_{\mathrm {ab}}\approx3$\,nm) is on the order of the spacing between the superconducting interlayer $c\approx0.8$\,nm) \cite{Tac89}. In thin films of high--\Tc\ cuprates grown by pulsed laser deposition, there is often a second local maximum at $B\parallel \mathrm {c}$ observed due to correlated defects which arise from the columnar growth in physical vapour deposition, PVD, methods. The absence of such a peak in our measurement as well as the absolute maximum values of \Jc\ in the region of $10^5$\,A/cm$^2$ is in agreement with the microstructure observed by TEM. At this point it should be emphasised that the phase formation during the ex--situ annealing process leads to a film growth with a high crystalline quality, because the film growth during the annealing process is more similar to single crystal growth than to typical PVD methods. 

It was shown for single band superconductors like YBa$_2$Cu$_3$O$_{7-\delta}$ that the angular dependence of $J_{\mathrm c}$ can be scaled using an effective field \Heff\,$=H \epsilon(\gamma_{\mathrm m},\Theta)$, where $\epsilon(\gamma_{\mathrm m},\Theta) = \sqrt{\sin^2{\Theta}+\gamma_{\mathrm m}^{-2}\cos^2{\Theta}}$, and $\gamma_{\mathrm m}^{2} = m_{\mathrm {c}}/m_{\mathrm {ab}}$ is the anisotropy of the effective electron masses. This scaling behaviour was theoretically shown by Blatter~et~al. \cite{Bla92}, and experimentally verified on clean YBa$_2$Cu$_3$O$_{7-\delta}$ thin films \cite{Civ04}, and on melt textured YBa$_2$Cu$_3$O$_{7-\delta}$ samples \cite{Bra93}. Since La--1111 is a multiband superconductor, different effective mass anisotropies can be expected for the different charge carriers, holes and electrons, in the individual bands, similar to MgB$_2$ \cite{Choi02,Ang02,Gur07}.

As shown in Fig.~\ref{fig:Criticalcurrents}b for four representative temperatures, \Jc\ of La--1111 can be scaled by \Heff\ in a wide angular range around $B\parallel \mathrm {c}$. From pinning force scaling, $F_{\mathrm p}=J_{\mathrm c}\cdot B$, $J_{\mathrm c}$($H_{\mathrm {eff}}$) can be recalculated and is plotted in Fig.~\ref{fig:Criticalcurrents}b (green lines). With these fit functions, which also resemble \Jc\ for $B\parallel \mathrm {c}$, the anisotropy curves due to random defects can be recalculated as shown in Fig.~\ref{fig:Criticalcurrents}a (green lines). Apart from intrinsic pinning due to the layered structure, the pinning centres -- typically point defects -- act isotropically and are randomly distributed. That \Jc\ scales in a large magnetic field range implies that the \Jc\ anisotropy essentially reflects the mass anisotropy of the effective charge carrier.

\begin{figure}[ht]
	\centering
		\includegraphics[width=1.0\columnwidth]{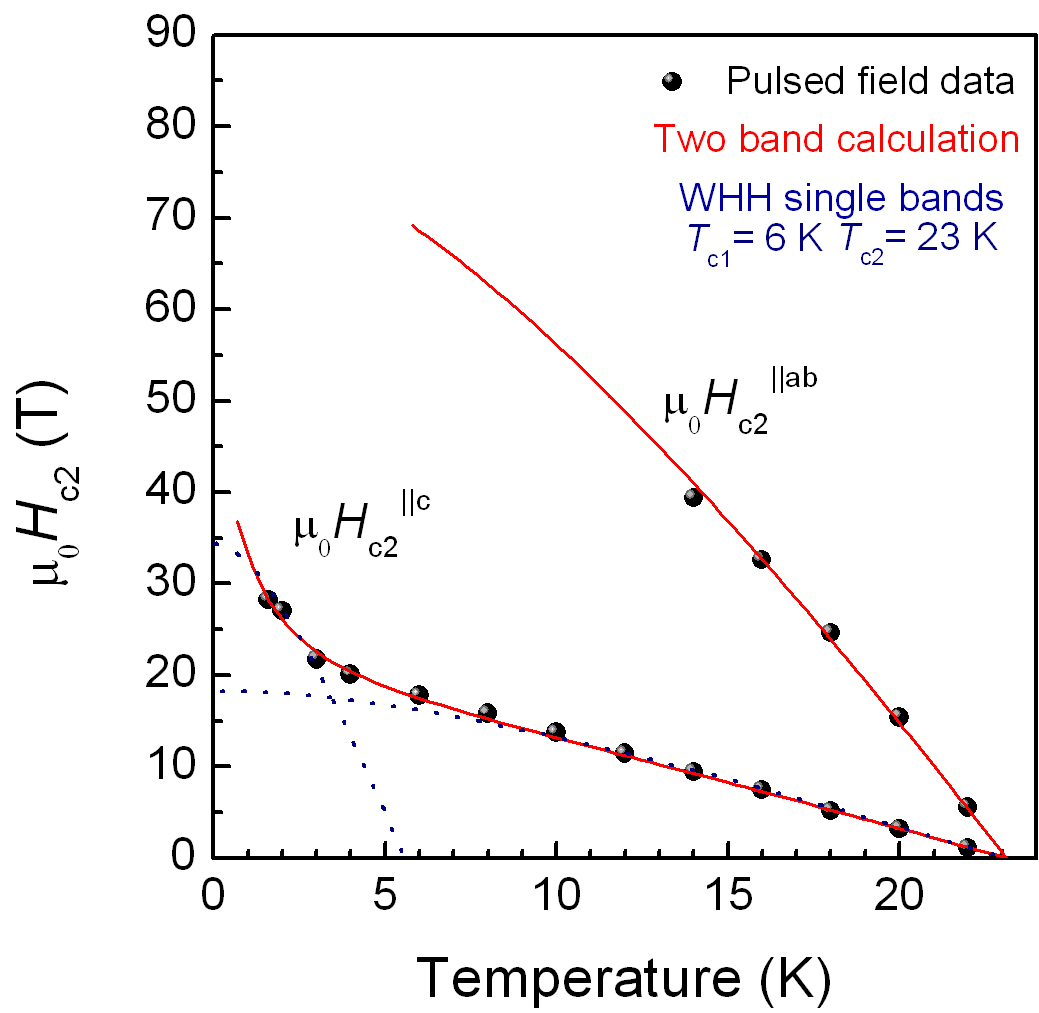}
	\caption{The magnetic phase diagram of La--1111. Due to multiband superconductivity, \Hcp\ shows a steep increase below 6\,K. Experimental data can be approximated using, as the simplest case, a two band model as described by Gurevich et al. \cite{Gur03,Gur07}(red lines). For weakly coupled bands, the physics can be described using two single bands. We assumed decoupled bands with different transition temperatures indicating this case (blue dotted line).}
	\label{fig:Phasendiagram}
\end{figure}

\begin{figure}[hb]
	\centering
		\includegraphics[width=0.66\columnwidth]{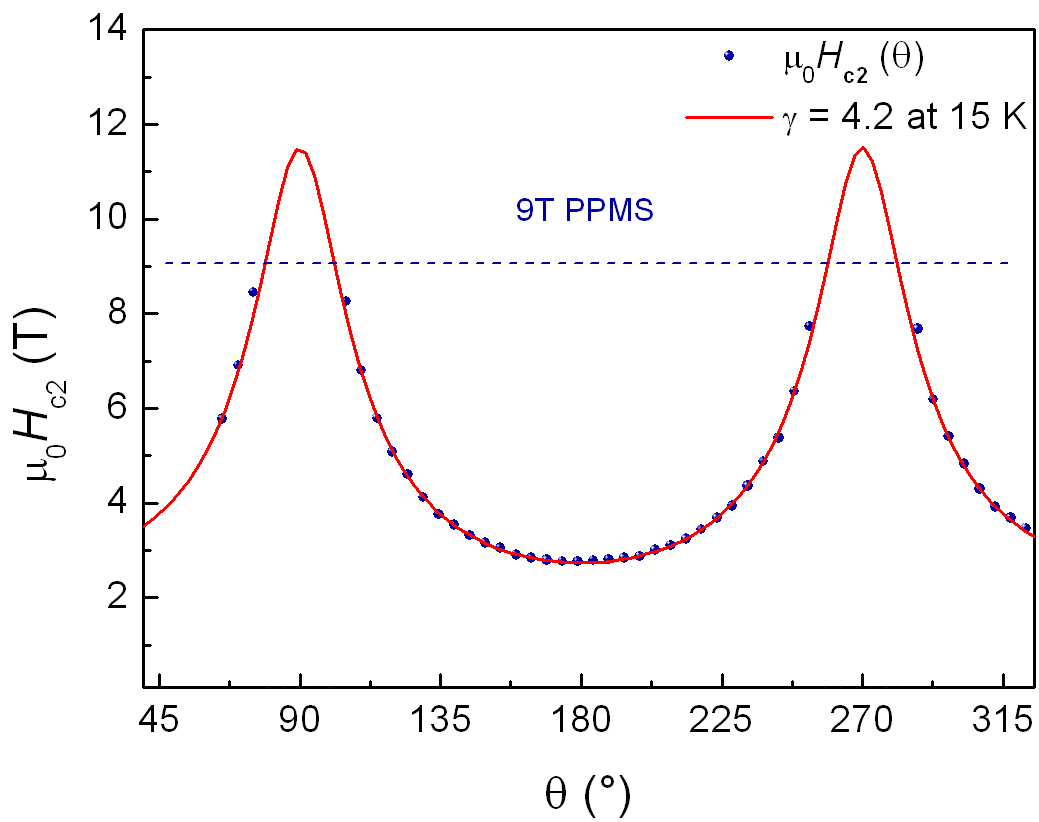}
	\caption{\Hc$(\theta)$ determined from angular dependent $R(B)$ measurements in a PPMS up to 9\,T using a resistance criterion of $R_{40\%}$. Experimental data can be described within the simple anisotropic Ginzburg Landau approximation $H_{\mathrm c 2}(\theta) = H_{\mathrm c 2}^{\parallel \mathrm {ab}}({\sin^2{\Theta}+\gamma_{H_{\mathrm c 2}}^{-2}\cos^2{\Theta}})^{-\frac{1}{2}}$ similar to the Blatter scaling. The evaluated anisotropy of the upper critical field $\gamma_{H_{\mathrm c 2}}$ is consistent with the data obtained from \Jc\ scaling and high field measurements as shown in Fig.~\ref{fig:Criticalcurrents}c.}
	\label{fig:GammaH}
\end{figure}

Evaluating the mass anisotropy from Blatter scaling shows a significant temperature dependence (Fig.~\ref{fig:Criticalcurrents}c). It ranges from $3.2$ at 2\,K to $4.2$ at 15\,K, the highest temperature we could access \Jc\ reliably at. At higher temperatures we were able to access $\gamma_{H_{c2}}={H_{\mathrm c 2}^{\parallel \mathrm {ab}}}/{H_{\mathrm c 2}^{\parallel \mathrm c}}$, which was calculated from measurements in pulsed (up to 42\,T Suppl:~Fig.~\ref{fig:Phasendiagram}) as well as static (up to 9\,T Suppl:~Fig.~\ref{fig:GammaH}) magnetic fields. The data at low temperatures, where \Hc\ is not accessible and at high temperatures, where \Jc\ is not accessible, join and overlap between 14 and 16 K indicating that the method of \Jc\ scaling probes the anisotropy of \Hc. As a result, we can describe the \Jc\ anisotropy within the theory of single band superconductors but with a temperature dependent mass anisotropy. This assumption is valid if the two bands are weakly coupled as discussed below. We can distinguish between an electronic band dominating the properties at low temperatures, which has an effective mass anisotropy of around $3$, and another band dominating at higher temperatures with an anisotropy of around $5$. It remains an open question which of the values of $\gamma_m \approx 3$ and $\gamma_m \approx 5$ can be related to electrons or holes.

\begin{figure}[htbp]
	\centering
		\includegraphics[width=0.85\columnwidth]{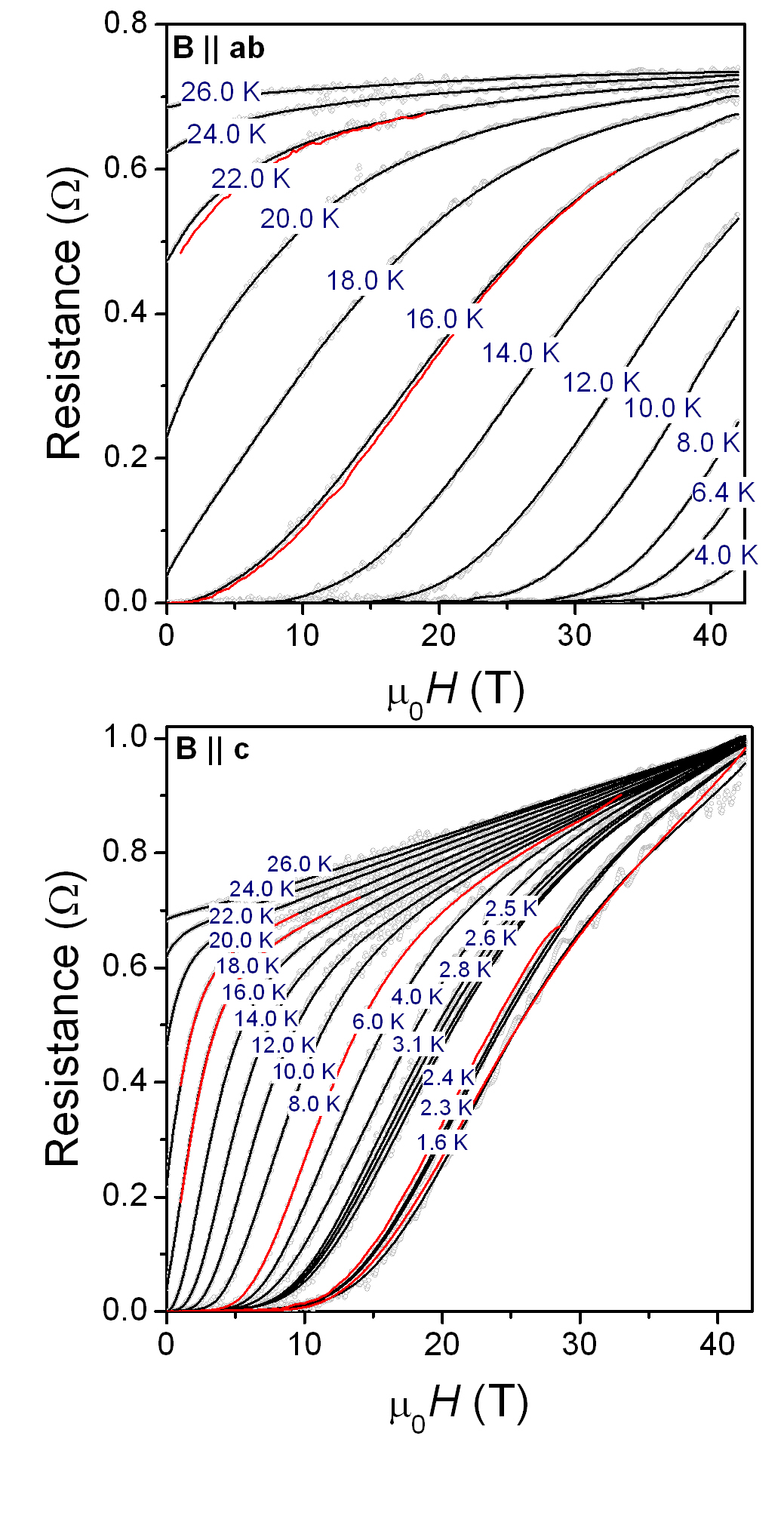}
	\caption{R(B) pulsed field measurements with B parallel and perpendicular to the \textit{c}--axis.
	Measurements in a pulsed magnetic field up to 42 T were realised using a cryostat equipped with a solenoid operated in pulsed mode. A detailed description of the pulsed field system can be found in [H. Krug et al., Physica B \textbf{294--295} 605 (2001)]. The resistance was measured using a standard four probe ac technique with a current amplitude of about 100\,$\mu$A and a frequency of 10\,kHz. Gold contacts deposited via standard pulsed laser deposition were used to provide low resistivity contacts to avoid possible heating effects during the pulse. Red lines represent pulses with different current amplitude and maximum field to exclude artifacts caused by the measurement conditions and possible heating of the sample. The \Rc\ criterion, where the resistance drops below 90\% of the resistance at 26\,K in zero field, was selected to evaluate the upper critical field at the corresponding temperature.}
	\label{fig:PulsedfeldSupp}
\end{figure}

Certainly, multiband superconductivity is manifested in the temperature dependent mass anisotropy as well as in the temperature dependence of the upper critical field. For the upper critical field parallel to the \textsl{c}--axis, \Hcp, we measured a steep upward curvature at low temperatures. The applied magnetic fields were not high enough to completely study the upper critical field perpendicular to the \textsl{c}--axis, \Hcs, in the low temperature region since $\mu_0$\Hcs\ exceeds 42\,T at 14\,K with a slope of {\small $\frac{d\:\mu_0H_{\mathrm c2}^{\parallel \mathrm {ab}}}{d\:T} \big| _{T_{\mathrm c}}$}$\approx-5$\,T/K near \Tc\ (Suppl:~Fig.~\ref{fig:PulsedfeldSupp},\ref{fig:Phasendiagram}). To describe the temperature dependence of \Hc\ we assume as the simplest case two bands, holes and electrons, using a model calculation originally developed by Gurevich \cite{Gur03} for MgB$_2$. The transition temperature of the weaker band is near 6\,K, where  a large increase of \Hcp\ is observed. Similar to MgB$_2$, the steep increase in \Hc\ at low temperatures may be explained by two weakly coupled bands, one of them in the dirty limit. In the case of two weakly coupled bands the physics can be described as two shunted single band superconductors \cite{Gur07}. 

\begin{figure}[htbp]
	\centering
		\includegraphics[width=0.90\columnwidth]{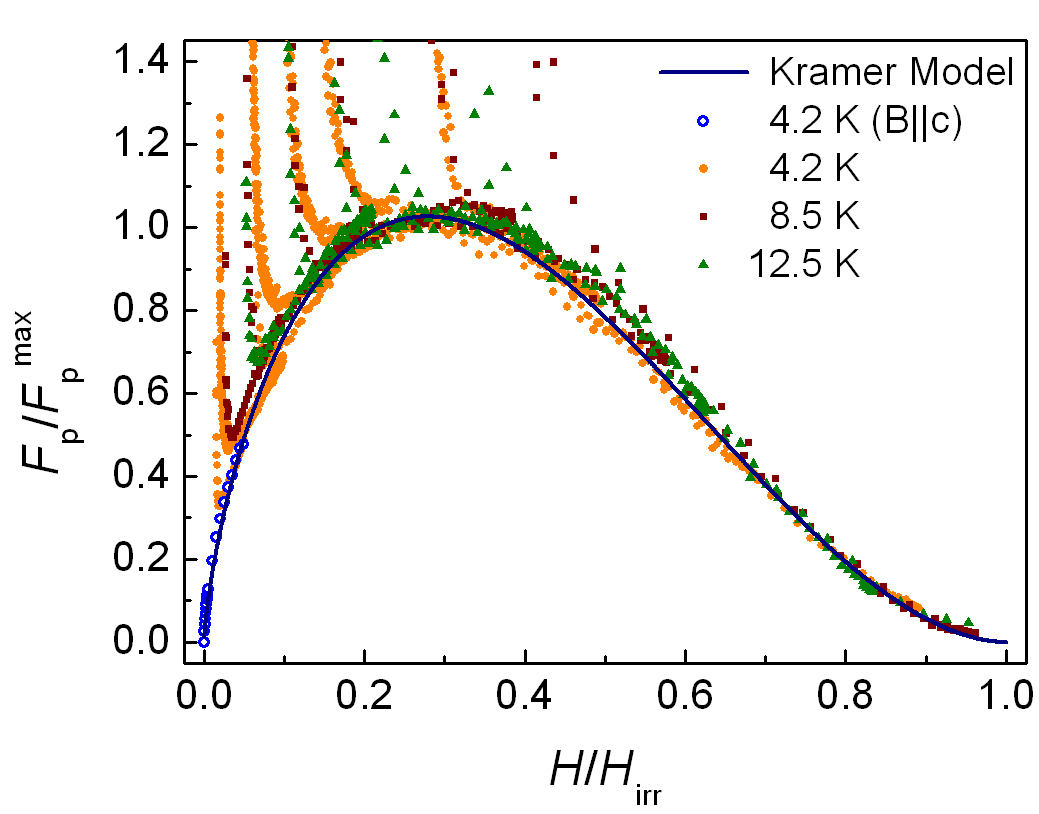}
	\caption{The normalised pinning force, $F_{\mathrm p}/F_{\mathrm {pmax}}$, over normalised magnetic field, ${H}/{H_{\mathrm {irr}}}$, scales in the entire temperature range. Hence, the pinning mechanism is temperature independent. Experimental data fit very well with the Kramer model ($p=0.7$ and $q=1.9$). Accordingly, \Jc\ is limited due to the shear breaking of the flux lines.}
	\label{fig:Pinningforce}
	\end{figure}
	
Without line defects and anisotropic precipitates, as confirmed by TEM and supported by the \Jc\ analysis, intrinsic and random pinning in oxypnictide superconductors can be studied. Moreover, iron pnictides are typical type--II superconductors similar to the high--\Tc\ cuprates with a short coherence length $\xi$ and a large penetration depth $\lambda$ resulting in a large $\kappa=\lambda/\xi$. The flux lines show plastic behaviour, especially in high magnetic fields, where the shear modulus, $C_{66}$, vanishes. For high--$\kappa$ superconductors the field dependence of $C_{66}$ can be approximated by $C_{66} \propto (\frac{H}{H_{\mathrm {irr}}})(1-\frac{H}{H_{\mathrm {irr}}})^2$ \cite{Bra86}.
In Fig.~\ref{fig:Pinningforce} the normalised pinning forces, $F_{\mathrm p}/F_{\mathrm {pmax}}$, at various temperatures are plotted over the normalised effective field ${H}/{H_{\mathrm {irr}}}$. For all temperatures, pinning forces scale for the region of random pinning, i.e. the pinning centres are randomly distributed and act isotropically. A functional relation is given by $F_{\mathrm p}/F_{\mathrm {pmax}}=K(\frac{H}{H_{\mathrm {irr}}})^p\:(1-\frac{H}{H_{\mathrm {irr}}})^q$
where $K$ is a proportionality constant, and $p$, and $q$ are fitting parameters. Best fits were obtained with $p=0.7$ and $q=1.9$ (blue line). This fits very well the model proposed by Kramer for shear breaking of the flux line lattice $F_{\mathrm p}/F_{\mathrm {pmax}}\propto(\frac{H}{H_{\mathrm {irr}}})^{\frac{1}{2}}\:(1-\frac{H}{H_{\mathrm {irr}}})^2$ \cite{Kra73}. As a result, the shear modulus $C_{66}$ dominates the pinning force in the entire range of the applied magnetic field \cite{Bra05}. In spite of multiband superconductivity the pinning force density can be scaled at all temperatures with the same fitting parameters. Hence, the pinning mechanism is temperature independent.

\begin{figure}[htbp]
	\centering
		\includegraphics[width=0.90\columnwidth]{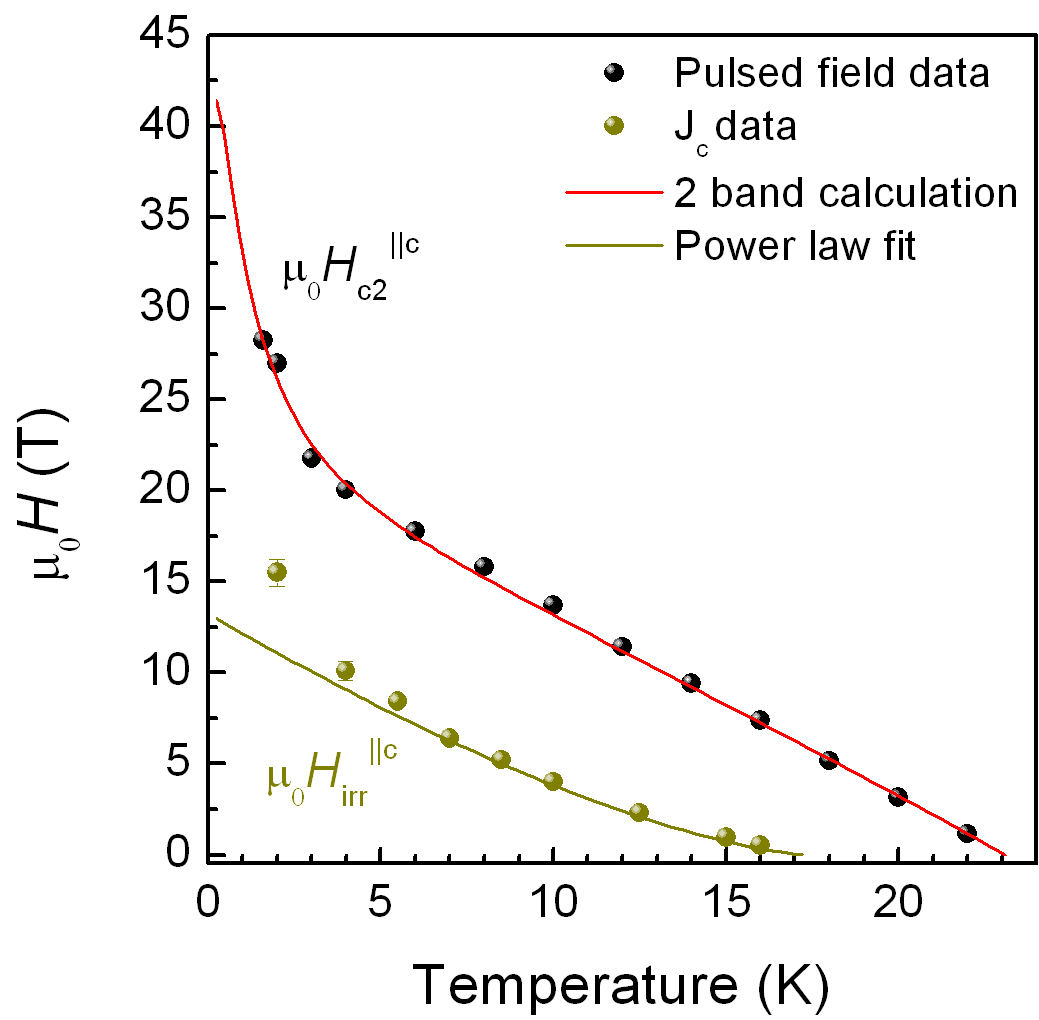}
	\caption{The part of the magnetic phase diagram for $B\parallel \mathrm c$ illustrates the steep increase of \Hc\ as well as \Hirr\ at low temperatures. The experimental data can be described assuming a two band superconductor.}
	\label{fig:HirrT}
	\end{figure}

The temperature dependence of the irreversibility field is included in the magnetic phase diagram in Fig.~\ref{fig:HirrT}. The data at very low temperatures were extrapolated from the pinning force scaling (Fig.~\ref{fig:Pinningforce}) where \Hirr\ is determined by the criterion of a vanishing pinning force. Additionally, we used Kramer plots ($J_{\mathrm c}^{\frac{1}{2}}B^{\frac{1}{4}} \ vs.\  B$) to extrapolate the irreversibility fields. A steep increase of \Hirrp\ can be seen at low temperatures in agreement with the steep increase of the upper critical field.

In summary, we have presented detailed critical current density measurements on an epitaxially grown superconducting \La\ thin film. The investigated sample reveals a clean \La\ layer which is free of correlated defects. We have demonstrated that the critical current densities of oxypnictides can be scaled using the Blatter approach with a temperature dependent anisotropy. Due to the general evidence of multiband superconductivity in the oxypnictides, the mass anisotropy of a distinct band is experimentally difficult to access. Nevertheless, since the \Jc\ anisotropy is scalable because of the temperature dependence of the upper critical field, we conclude that \La\ has two major weakly coupled bands. This enables the evaluation of the mass anisotropies in the limit of low or high temperatures, respectively. We were able to explore the upper critical field parallel to the c--axis in a large temperature range. The steep increase of the upper critical field as well as of the irreversibility field at low temperatures suggests the responsible band is in the dirty limit due to point defects. Accordingly, \La\ shows a strong analogy to MgB$_2$. Combining \Hc\ data and the scaling of \Jc\ we were able to deduce the temperature dependence of the effective mass anisotropy, which can not be measured directly due to the high upper critical fields of the oxypnictides.

\textit{Acknowledgements.} 
This work was partially funded by the EU-FP6 Research Project NanoEngineered Superconductors for Power Applications (NESPA) no. MRTN-CT-2006-035619 and by the German Research Foundation (DFG) under project HA 5934/1-1. We thank O.~G.~Schmidt, S.~Baunack and C.~Deneke for help with the sample preparation in the FIB as well as B.~Rellinghaus and D.~Pohl for assistance with the interpretation of the TEM data. Finally, we thank J.~Werner and M.~Langer for the target preparation.

\clearpage


\begin{figure}[htbp]
	\centering
		\includegraphics[width=0.9\columnwidth]{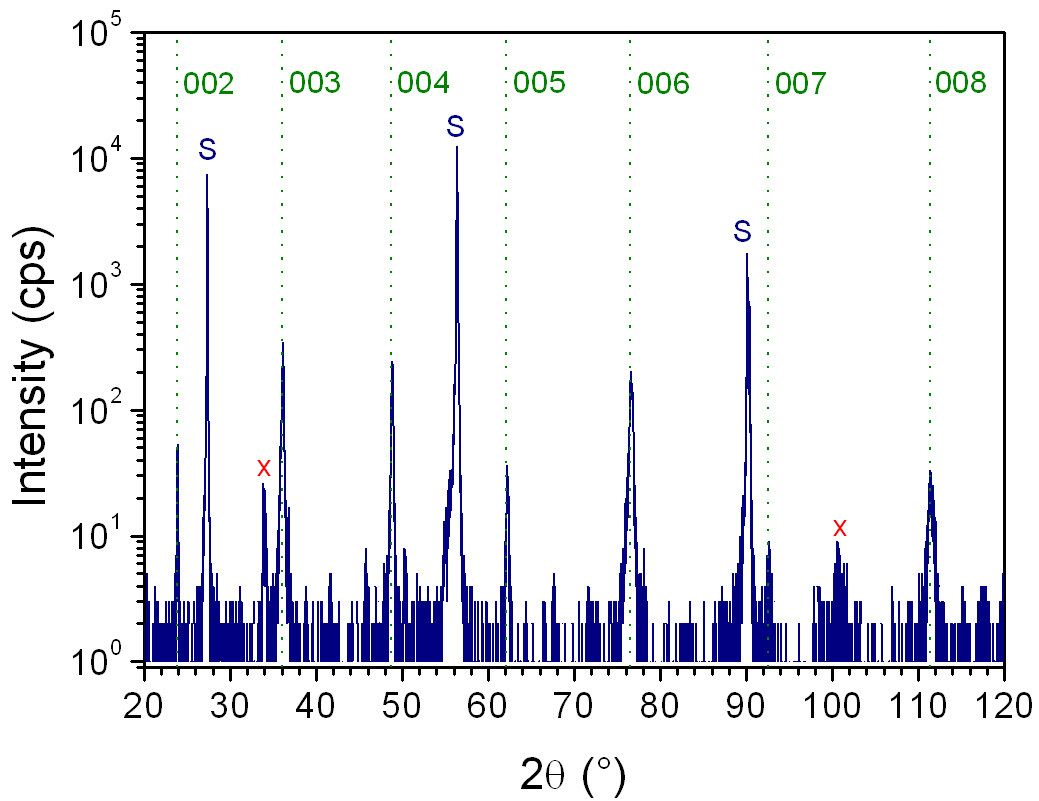}
	\caption{X--ray diffraction pattern for a La--1111 thin film with the indexed (00l) peaks of the superconducting \La\ phase. The substrate is marked with "S" and impurities with "x". The signals for impurities come exclusively from the polycrystalline surface layer and are not present in the superconducting La--1111 layer.}
	\label{fig:xray}
\end{figure}

\begin{figure}[htbp]
	\centering
		\includegraphics[width=0.9\columnwidth]{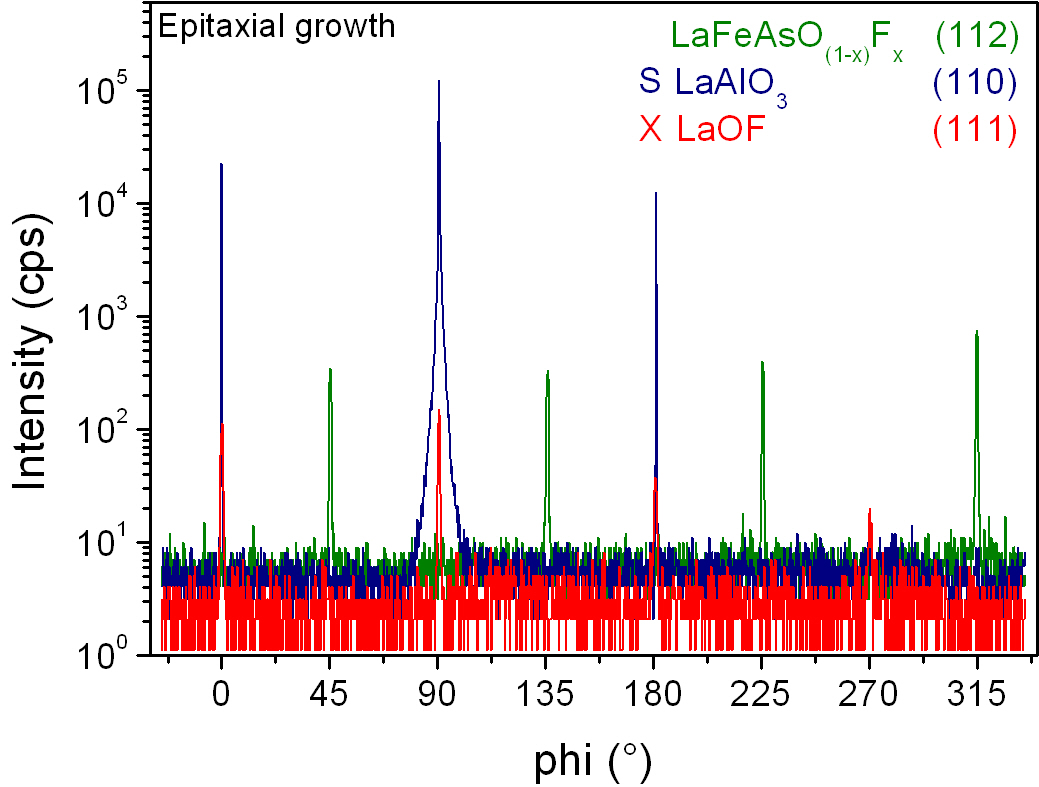}
	\caption{X--ray phi--scans proving epitaxial growth and reinforcing the results from TEM in Fig.~\ref{fig:HRTEM}c. Different peak heights are due to slightly tilted sample mounting.}
	\label{fig:Texture}
\end{figure}

\begin{figure}[htbp]
	\centering
		\includegraphics[width=1\columnwidth]{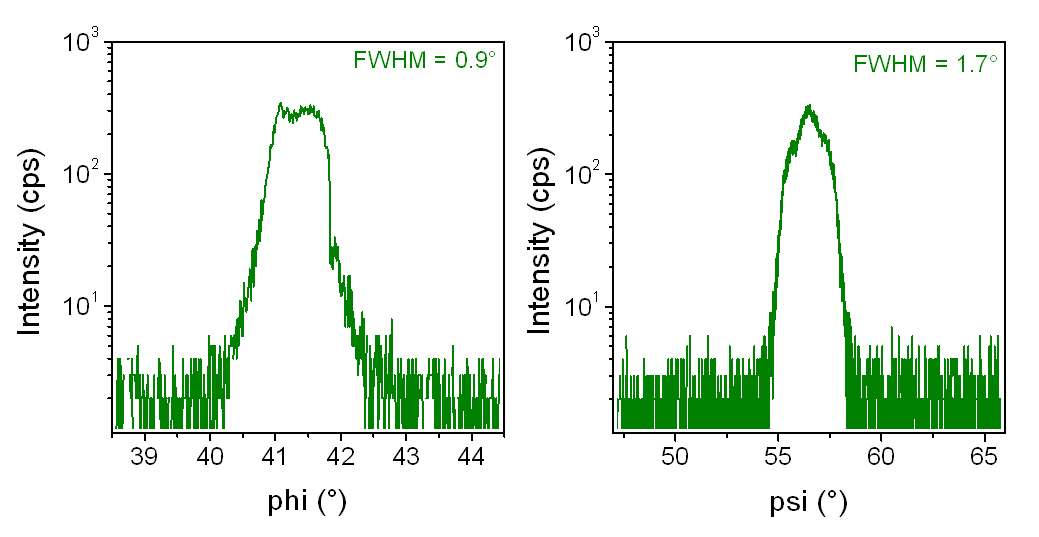}
	\caption{A phi-- and psi--scan of the \La\ (112) pole reveals a very sharp in--plane and out--of--plane orientation distribution with a full width at half maximum value of 0.9° and 1.7°, respectively.}
	\label{fig:Texturedetails}
\end{figure}

\clearpage

\newpage

\newpage

\newpage

\end{document}